\documentclass[twocolumn,showpacs,preprintnumbers,amsmath,amssymb,floatfix]{revtex4}
\usepackage{graphicx}

\begin{document}


\title{Inelastic electron tunneling via molecular vibrations in single-molecule transistors}

\author{L.H. Yu$^{1}$, Z.K. Keane$^{1}$, J.W. Ciszek$^{2}$, L. Cheng$^{2}$, M.P. Stewart$^{2}$, J.M. Tour$^{2}$, D. Natelson$^{1,3}$}

\affiliation{$^{1}$ Department of Physics and Astronomy, $^{2}$ Department of Chemistry and Center for Nanoscale Science and Technology, $^{3}$ Department of Electrical and Computer Engineering, Rice University, 6100 Main St., Houston, TX 77005}

\date{\today}

\begin{abstract}

In single-molecule transistors, we observe inelastic cotunneling
features that correspond energetically to vibrational excitations of
the molecule, as determined by Raman and infrared spectroscopy.  This
is a form of inelastic electron tunneling spectroscopy of single
molecules, with the transistor geometry allowing in-situ tuning of the
electronic states via a gate electrode.  The vibrational features shift
and change shape as the electronic levels are tuned near resonance,
indicating significant modification of the vibrational states.  When
the molecule contains an unpaired electron, we also observe
vibrational satellite features around the Kondo resonance.

\end{abstract}

\maketitle

\newpage

Electron tunneling is widespread throughout chemistry and condensed
matter physics.  Electron transfer through molecules by nonresonant
tunneling has long been known\cite{Marcus56JCP,McConnell61JCP}, and
tunneling electrons can interact inelastically with molecules,
exciting vibrational modes.  This is the basis of inelastic electron
tunneling spectroscopy (IETS)\cite{JaklevicetAl66PRL,HippsetAl93JPC},
recently refined to probe groups of molecules via crossed
wires\cite{Gregory90PRL,ZimmermanetAl01RSI,KushmericketAl04NL} and
nanopores\cite{WangetAl04NL}, and single molecules via scanning
tunneling microscopy (STM)\cite{StipeetAl98Science}.  Similar
inelastic, nonresonant tunneling occurs in single-electron transistors
(SETs)\cite{Averinbook,DeFranceschietAl01PRL}, three-terminal devices
in which a gate electrode allows in-situ adjustment of the energy
levels.  To date, this tuning has not been possible in the chemical
systems examined by IETS.

In this Letter we report inelastic cotunneling processes in
single-molecule transistors
(SMTs)\cite{ParketAl00Nature,ParketAl02Nature,LiangetAl02Nature,YuetAl04NL},
and identify them with vibrational excitations of the molecules, as
determined by Raman and infrared spectroscopy.  Tuning electronic
levels near resonance reveals shifts in inelastic lineshapes and peak
positions, suggesting significantly modified electron-vibrational
coupling in this region.  The vibrational features persist in the
Kondo regime, indicating a complicated conduction process involving
vibrational excitations of a many-body electronic system.

\begin{figure}[h!]
\begin{center}
\includegraphics[clip, width=8cm]{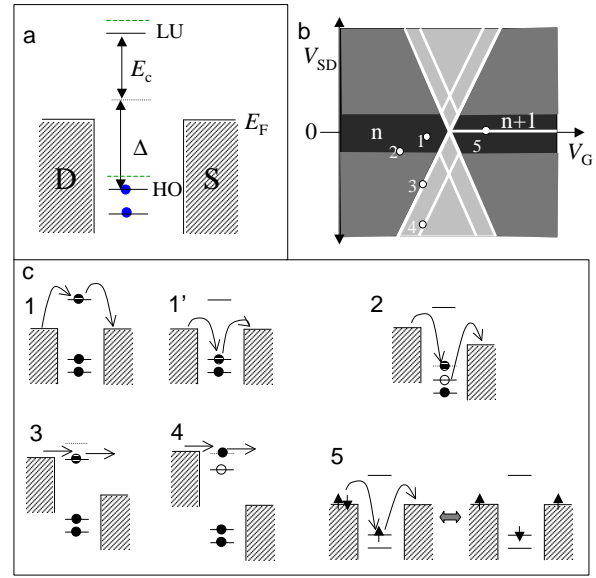}
\end{center}
\vspace{-5mm}
\caption{\small (a) Energetics of single-electron transistor, with $\Delta \equiv$ single-particle level spacing, $E_{c}\equiv$ charging energy.  Dashed levels indicate manifold of vibrational excitations in molecular devices. (b) Map of $\partial I_{D}/\partial V_{SD}$ (brightness) of such an SET, vs. source-drain bias $V_{SD}$ and gate voltage $V_{G}$, showing one Coulomb transition from $n$ to $n+1$ electrons on the island. (c) Conduction processes corresponding to the numbered points in (b): 1 = elastic cotunneling; 2 = inelastic cotunneling; 3 = resonant conduction; 4 = resonant inelastic conduction; 5 = Kondo resonant conduction. }
\label{fig1}
\vspace{-5mm}
\end{figure}

A confined electronic system (``island") coupled by tunneling to
source and drain electrodes is shown schematically in
Fig.~\ref{fig1}a.  The energy to promote an electron from the highest
occupied (HO) to the lowest unoccupied (LU) orbital is the
single-particle level spacing, $\Delta$.  Ignoring spin, adding an
electron to the island also requires additional Coulomb "charging
energy", $E_{c}$, often approximated in SETs by the charging energy of
a classical capacitor.  In molecules, a manifold of vibrational
excitations is associated with each electronic state.  In a
three-terminal device, the island levels may be shifted via a gate
potential.  Fig.~\ref{fig1}b maps the differential conductance,
$\partial I_{D}/\partial V_{SD}$, of a generic SET as a function of
source-drain bias, $V_{SD}$, and gate voltage, $V_{G}$, while
Fig.~\ref{fig1}c shows possible electronic transport mechanisms.  For
low biases (Fig.~\ref{fig1}c, 1,1') the average number of electrons on
the island is fixed; transport is suppressed (Coulomb blockade) and
can only occur by higher order tunneling through virtual states.  An
example of this in chemical electron transfer is
``superexchange"\cite{McConnell61JCP}, and in SETs such processes are
called ``elastic cotunneling"\cite{Averinbook}.  At higher biases in
the blockaded regime, ``inelastic cotunneling" via an excited virtual
state (Fig.~\ref{fig1}c, 2) is possible.  For an excitation of energy
$E^{*}$, the opening of the inelastic channel results in a feature in
$\partial^{2}I_{D}/\partial V_{SD}^{2}$ at $eV_{SD} = E^{*}$.
Inelastic cotunneling via electronically excited states has been seen
in semiconductor\cite{DeFranceschietAl01PRL} and carbon nanotube
SETs\cite{LiangetAl02PRL}.  Inelastic cotunneling via vibrationally
excited molecules is responsible for conventional
IETS\cite{TroisietAl03JCP}, but has not been studied in three-terminal
devices.  IETS lineshapes are predicted to vary significantly
depending on the energetics of the virtual
states\cite{GalperinetAl04condmat}, and can be peaks, dips, or
intermediate structures in $\partial^{2}I_{D}/\partial V_{SD}^{2}$.
At still higher source-drain biases (Fig.~\ref{fig1}c, 3) Coulomb
blockade is lifted leading to significant resonant conduction, while
at still higher biases (Fig.~\ref{fig1}c, 4) additional resonant
conduction occurs when $eV_{SD}$ is sufficient to leave the island in
an
electronically\cite{JohnsonetAl92PRL,StewartetAl97Science,KouwenhovenetAl97Science}
or
vibrationally\cite{ParketAl00Nature,ParketAl02Nature,YuetAl04NL,QiuetAl04PRL}
excited state.

When the effects of unpaired spins are included, the Kondo resonance
becomes a possible conduction mechanism, and is detected as a sharp
conductance peak near zero
bias\cite{GoldhaberGordonetAl98Nature,CronenwettetAl98Science} at
temperatures low compared to a characteristic energy scale, $k_{\rm
B}T_{\rm K}$.  The maximum peak conductance possible is $2e^2/h$, for
a system with perfectly symmetric coupling between the island and the
source and drain electrodes.  Kondo physics has been observed recently
in single-molecule
transistors\cite{ParketAl02Nature,LiangetAl02Nature,YuetAl04NL}.

\begin{figure}[h!]
\begin{center}
\includegraphics[clip, width=8cm]{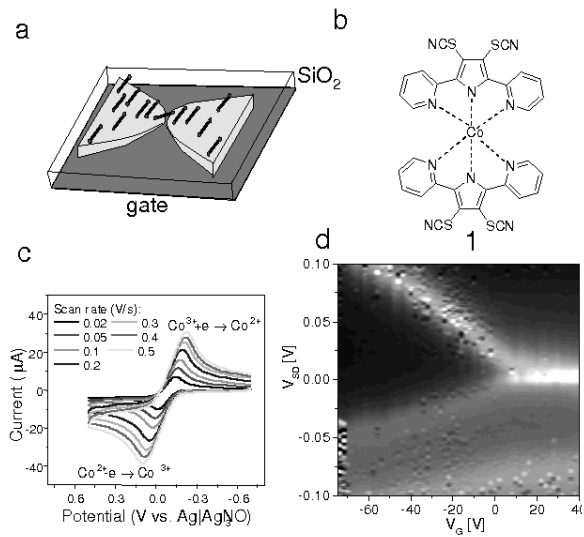}
\end{center}
\vspace{-5mm}
\caption{\small (a) Diagram of typical single-molecule transistor fabricated
by electromigration.  (b) Structure of {\bf 1}, the compound of interest. (c) Cyclic voltammograms showing reversible change in Co ion charge state\protect{\cite{supp}}.  (d) Differential conductance of a transistor containing {\bf 1}, at 5~K.  White corresponds to $\partial I_{D}/\partial V_{SD} = 3 \times 10^{-6}$~S.  The zero-bias peak in the righthand charge state is characteristic of Kondo resonant conduction.}
\label{fig2}
\vspace{-5mm}
\end{figure}

Single-molecule transistors open the possibility of examining
inelastic cotunneling in individual, tunable chemical systems, in both
the blockaded and Kondo regimes.  We fabricate SMTs (Fig.~\ref{fig2}a) using an
electromigration method that has been described extensively\cite{ParketAl00Nature,ParketAl02Nature,LiangetAl02Nature,YuetAl04NL}.  The source and drain electrodes are 15~nm Au films with 1~nm Ti
adhesion layers, prepared by electron beam lithography, e-beam
evaporation, liftoff, and oxygen plasma cleaning.  The gate oxide is
200~nm SiO$_{2}$, with a degenerately doped $p+$ Si (100) wafer as the
underlying gate electrode.  While this thickness of oxide guarantees
relatively weak gate couplings due to simple geometric considerations,
we routinely obtain excellent device reliability up to gate fields of
$5 \times 10^{8}$V/m with negligible leakage.

The starting molecule, {\bf 1}, (Fig.~\ref{fig2}b) comprises a single
transition metal ion (Co$^{2+}$) coordinated by conjugated ligands;
the valence state of the ion may be controlled electrochemically
(Fig.~\ref{fig2}c).  The structure of {\bf 1} has been verified by
x-ray crystallography.  Compound {\bf 1} undergoes loss of the (CN)
moieties upon assembly on gold in tetrahydrofuran (THF), yielding the
corresponding di- or tri-thiolate, which is bound covalently to
surface Au atoms\cite{CiszeketAl04JACS}.  Both {\bf 1} and its
self-assembly have been characterized extensively by x-ray
photoemission, ellipsometry, and electrochemical methods\cite{supp}.

A 2mM solution of {\bf 1} in THF is allowed to self-assemble on the
electrode sets for 48 hours.  The chip is then rinsed in THF, dried in
a nitrogen stream, and placed in a variable temperature vacuum probe
station (Desert Cryogenics).  This cryostat does not have magnetic
field capabilities, but does permit characterization of many junctions
at once.  The junctions are partially broken to the k$\Omega$ level by
electromigration at room temperature, cooled to liquid helium
temperatures, and broken by further electromigration into separate
source and drain electrodes.

At a variety of temperatures, we have measured $I_{D}$ as a function
of $V_{SD}$ at $V_{G}$ from -100~V to +100~V using a semiconductor
parameter analyzer (HP4145B), with the source grounded and the drain
electrode swept.  We compute the differential conductance $\partial
I_{D}/\partial V_{SD}$ and $\partial^{2}I_{D}/\partial V_{SD}^{2}$ as
a function of $V_{SD}$ and $V_{G}$ by numerical differentiation.  Spot
comparisons between this approach and lock-in techniques show
excellent agreement\cite{supp}.  To avoid artifacts we identify features in
$\partial^{2}I_{D}/\partial V_{SD}^{2}$ by comparing multiple data
sets taken at the same or nearby gate voltages.

We have examined 407 electrode pairs on 10 separate substrates, with
statistics similar to those in previous
investigations\cite{YuetAl04NL}.  Of the electrode pairs examined we
found 57 devices with no detectable current (electrodes too far apart
for measurable conduction); 166 with linear, nongateable
current-voltage characteristics (likely no molecule present at the
junction); 108 with nonlinear but ungateable current-voltage
characteristics (either molecules or metal nanoparticles with
negligible gate coupling); and 76 nonlinear but significantly gateable
current-voltage characteristics.  From this last group, four were
identified as single-electron devices based on unintentionally
produced metal nanoparticles.  This identification was based on the
observation of many regularly spaced Coulomb blockade regions with
typical electron addition energies less than 50~meV.  In working
devices, bias sweeps were restricted to $|V_{SD}| < 200$~mV to
minimize the chances of current-induced irreversible changes.  Device
stability is poor at high current densities and temperatures
significantly above 4.2 K, likely due to the atomic diffusion of the
Au electrode material.

There are 27 devices that cleanly display a single Coulomb degeneracy
point, the vast majority of which show a resonance at zero $V_{SD}$ in
one charge state identified as a Kondo resonance, as in
Fig.~\ref{fig2}d.  Control devices with electrodes exposed to THF
without {\bf 1} never display such conductance properties.  As can be seen
from the edges of the Coulomb blockade region, the electron addition
energy for the device in Fig.~\ref{fig2}d exceeds 100~meV; this is
typical.  The Kondo properties are very similar to those reported in
earlier SMTs based on Co$^{2+}$-containing
complexes\cite{ParketAl02Nature}.  Typical Kondo temperatures as
inferred from the temperature dependence of the Kondo resonance height
and the low temperature width of the resonance are $\sim$40~K.  The
low temperature limit of the resonance peak height is often reduced
from the theoretical maximum value of $2e^{2}/h$, indicating
asymmetric coupling of the molecule to the source and drain
electrodes.  A full discussion of Kondo physics in {\bf 1} and related
molecules will be reported elsewhere.

\begin{figure}[h!]
\begin{center}
\includegraphics[clip, width=8cm]{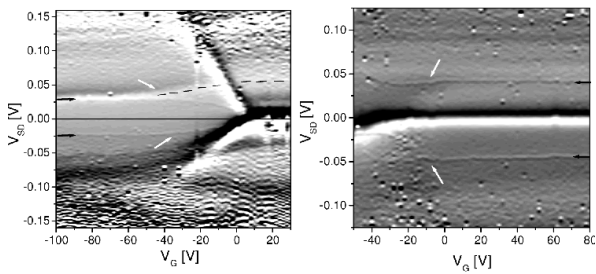}
\end{center}
\vspace{-5mm}
\caption{\small Maps of $\partial^{2} I_{D}/\partial V_{SD}^{2}$ as a function of $V_{SD}$ and $V_{G}$ at 5~K for two devices.  Smoothing window in $V_{SD}$ is 5~mV.  Brightness scales are $-8 \times 10^{-5}$~A/V (black) to $3 \times 10^{-5}$~A/V (white), and $-2 \times 10^{-5}$~A/V (black) to $2 \times 10^{-5}$~A/V, respectively.  The zero-bias features correspond to Kondo peaks in $\partial I_{D}/\partial V_{SD}$.  Prominent inelastic features are indicated by black arrows.  In both devices, when the inelastic features approach the boundaries of the Coulomb blockade region, these levels shift and alter lineshape (white arrows).  Black dashed line in left map traces an inelastic feature across across the boundary and into the Kondo regime.}
\label{fig3}
\vspace{-5mm}
\end{figure}

In 12 devices, the conductance in the classically blockaded region
and/or outside the Kondo resonance is large enough to allow clean
measurements of $\partial^{2}I_{D}/\partial V_{SD}^{2}$.  In
Figs.~\ref{fig3}, we show maps of this quantity as a function of
$V_{SD}$ and $V_{G}$ in two different devices at 5~K.  We have
indicated two prominent features within the blockaded (Kondo) regime
with black arrows.  Features in $\partial^{2}I_{D}/\partial
V_{SD}^{2}$ of opposite sign are symmetrically located around zero
source-drain bias, consistent with features expected from inelastic
tunneling.  Some asymmetry in shape is unsurprising, given that the
low peak conductance in the Kondo regime for these devices indicates
significantly asymmetric coupling of the molecule to the leads.

The $\partial^{2}I_{D}/\partial V_{SD}^{2}$ features in the blockaded
region occur at essentially constant values of $V_{SD}$ until $V_{G}$
is varied such that the feature approaches the edge of the blockaded
region.  This constancy in $V_{G}$ has been identified as a feature of
inelastic cotunneling in semiconductor\cite{DeFranceschietAl01PRL} and
nanotube\cite{LiangetAl02PRL} single-electron devices.  The changes in
these features as the boundaries of the Coulomb stability region are
crossed indicate that the inelastic processes are native to the SMT
itself, and not due to some parallel conduction channel.  The
inelastic modes occur at energies low compared to $\Delta$ ($\sim$100s of meV),
implying that the modes being excited are unlikely to be electronic.
By examining the $\partial^{2}I_{D}/\partial V_{SD}^{2}$ data for
features at fixed $V_{SD}$ as a function of $V_{G}$, we identified a
total of 43 candidate inelastic features at $V_{SD} < 120$~meV,
ranging in apparent width from 5~mV to 20~mV at $\sim$5~K.  The
narrowest inelastic features clearly show a broadening as the
temperature is elevated above 20~K.  Further studies will seek to
compare this broadening with standard theoretical treatments of
linewidths in IETS.

\begin{figure}[h!]
\begin{center}
\includegraphics[clip, width=8cm]{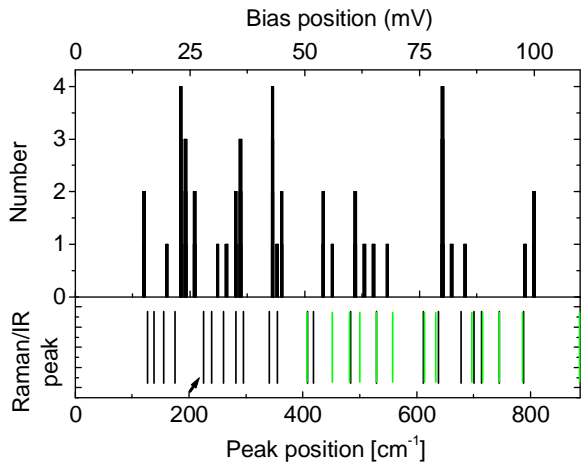}
\end{center}
\vspace{-5mm}
\caption{\small (top) Histogram of $|V_{SD}|$ positions of 43 features in $\partial I_{D}^{2}/\partial V_{SD}^{2}$ that are persistent at constant $V_{SD}$ over at least a 10~V gate voltage range, for the 12 samples discussed.  Bin size is 1~meV.  Width of actual features is at least 5~meV, with position indicating feature center. (bottom) Raman (black) and IR (gray) peak positions taken of {\bf 1} in the solid state at room temperature.  The indicated Raman peak near 28~meV is only present in films of {\bf 1} self-assembled on Au electrodes.  }
\label{fig4}
\vspace{-5mm}
\end{figure}

Fig.~\ref{fig4} shows a histogram (bin = 1 mV) of the $V_{SD}$
positions of those features, from detailed examination of
$\partial^{2}I_{D}/\partial V_{SD}^{2}$ vs. $V_{SD}$ data.  The lower
panel shows Raman (Stokes) peak positions (data taken on a powder of
{\bf 1}) and IR absorption peak positions of {\bf 1} (in pellet form,
blended with KBr)\cite{supp}.  The correlations between the spectra and
inelastic features in $\partial^{2}I_{D}/\partial V_{SD}^{2}$ confirm
that vibrational inelastic cotunneling processes are at work in the
molecular transistors.  This is consistent with the observation of
IETS in single molecules by STM\cite{StipeetAl98Science}.  The fact
that different SMTs exhibit different subsets of features is a natural
consequence of the sensitivity of IETS to the nanoscale structure of
the molecular junction\cite{ChenetAl04condmat}.  Comparisons with
Raman of {\bf 1} and related compounds (e.g. ligands without -SCN;
ligands without Co$^{2+}$) suggest tentative peak assignments.  The
peaks near 36~meV and 55~meV are likely the Co-N
stretch\cite{Socratesbook}, while that near 44~meV is tentatively a
-SCN deformation\cite{Socratesbook} (implying that not every S is
bonded to an Au surface atom, consistent with x-ray photoemission
data\cite{supp}).  The 24~meV peak is tentatively the Au-S
bond\cite{KatoetAl02JPCB}; this claim is further supported by the
appearance of a $\sim$28~meV Raman peak in films of {\bf 1} assembled
on Au electrodes, while such a peak is absent in Raman spectra of both
unassembled {\bf 1} and bare electrodes.

As indicated by white arrows in Fig.~\ref{fig3}, inelastic features
shift and change near the boundary of the Coulomb blockade region.
These changes as a function of $V_{G}$ are qualitatively consistent
with recent theoretical expectations\cite{GalperinetAl04condmat},
though detailed modelling would be required for quantitative
comparisons.  The energy of the virtual cotunneling states is shifted
by $V_{G}$, altering the (complex) amplitude for that process.  Changing
the cotunneling amplitude compared to quasi-elastic (with the virtual
excitation and readsorption of a vibrational quantum) or direct
source-drain tunneling is predicted to severely alter lineshapes.
These effects cannot be examined in two-terminal tunneling structures,
which lack the ability to shift the relevant virtual states.  We note
that occasionally inelastic features shift significantly in energy as
the blockade boundary is approached (e.g. the 24~meV features in
Fig.~\ref{fig3}(left)).  This would be consistent with a modified (dressed)
electron-vibrational coupling near the electronic resonance.

We observe that the IETS features persist into the Kondo regime in the
form of satellite features paralleling the zero-bias Kondo resonance.
Often specific features may be traced from the blockaded region into
the Kondo region, an example of which is indicated by the dashed line
in Fig.~\ref{fig3}(left).  Through the transition the inelastic
feature lineshapes change, as do their intensities.  Such inelastic
satellites paralleling the Kondo resonance have been suggested
previously\cite{LiangetAl02Nature,YuetAl04NL}, and the consistency in
feature position in our data lend strong support to this idea.  Recent
experiments in semiconductor devices\cite{KoganetAl04Science} have
demonstrated the existence of satellite Kondo peaks when the Kondo
system can interact inelastically with photons of a well defined
energy.  In the molecular transistor case, the inelastic exchange
occurs with the vibrational quanta of the molecule, demonstrating a
quantum mechanically coherent coupling between the electronic
many-body Kondo state, and the mechanical resonances of the molecule.
These Kondo measurements illustrate how the tunability of
single-molecule transistors permits the examination of vibrational
processes as a function of the energetics of the electronic levels, a
study not possible in standard two-terminal devices.

The authors thank A. Nitzan and M. Di Ventra for useful discussions.
DN acknowledges financial support from the Research Corporation, the
Robert A. Welch Foundation, the David and Lucille Packard Foundation,
an Alfred P. Sloan Foundation Fellowship, and NSF award DMR-0347253.
JMT acknowledges support from DARPA and the ONR.



\end{document}